\documentclass[prb,twocolumn,preprintnumbers,amsmath,amssymb]{revtex4}

\usepackage{graphicx}
\usepackage{amssymb,amsmath}
\usepackage{dcolumn}
\usepackage{bm}
\usepackage{color}
\usepackage{enumerate}

\begin{document}

\title{Spin-orbit coupled cold exciton condensates}

\author{O. Kyriienko}
\affiliation{Science Institute, University of Iceland, Dunhagi-3,
IS-107, Reykjavik, Iceland}

\author{E. B. Magnusson}
\affiliation{Science Institute, University of Iceland, Dunhagi-3,
IS-107, Reykjavik, Iceland}

\author{I. A. Shelykh}
\affiliation{Science Institute, University of Iceland, Dunhagi-3,
IS-107, Reykjavik, Iceland}
\affiliation{Division of Physics and Applied Physics, Nanyang Technological University 637371, Singapore}

\date{\today}

\begin{abstract}
We analyze theoretically the dynamics of degenerate condensate of cold indirect excitons.  We account for both linear spin dependent terms arising from spin-orbit interaction of Rashba and Dresselhaus types and non-linear terms transforming a pair of bright excitons into a pair of dark ones. We show that both terms should lead to the qualitative changes in the dynamics of cold exciton droplets in the real space and time.
\end{abstract}

\maketitle

\section{Introduction}
Collective phenomena lie behind many remarkable effects in physics. One of their famous manifestations is Bose-Einstein condensation,\cite{ReviewBEC} which occurs if a system of Bose particles is cooled down beyond the critical temperature $T_c$ which strongly depends on properties of the individual particles, in particular their effective mass. For systems of cold atoms, where condensing particles are very heavy, $T_c$ lies in the nano-Kelvin regime, which rules out any possibility of the practical implementation of this phenomenon.

On the other hand in the field of condensed matter physics various candidates were proposed for the realisation of BEC with critical temperatures orders of magnitudes higher then those of cold atoms.\cite{SnokeBook} The formation of exciton condensates in bulk semiconductors was theoretically predicted more than 40 years ago,\cite{Keldysh} but appeared to be difficult to realize experimentally. Since then, other solid-state candidates were proposed for achieving high-temperature BEC, including Quantum Hall bilayers,\cite{Eisenstein} magnons,\cite{Democritov} cavity exciton-polaritons \cite{KasprzakNature,Balili,BaumbergPRL2008} and indirect excitons.\cite{Timofeev,Butov1} The latter system is in focus of the present paper.

Spatially indirect exciton is a bound state of an electron and a hole localized in coupled parallel 2D layers. Electron and hole wave functions show a very small overlap and consequently indirect excitons have a long lifetime as compare to ordinary excitons. They behave as metastable particles which allows their cooling beyond the temperature of a quantum degeneracy.\cite{Butov2,SnokeScience}

The indirect excitons have been widely studied both experimentally and theoretically in recent years. Superfluid behavior of a system of indirect excitons was predicted by Lozovik and Yudson more than 30 years ago \cite{Lozovik} and subsequent theoretical \cite{Shevchenko,Zhu,Berman} and experimental \cite{Larionov,Snoke2002,Butov2002} studies have suggested that this should be manifested in a series of remarkable effects, including persistent currents, Josephson-related phenomena and spontaneous pattern formation in the real space.

Surprisingly, most of the works dedicated to indirect excitons have neglected their spin structure. On the other hand, it became clear in recent years that the account of the spinor nature of the condensing bosons can lead to new qualitative phenomena. For cavity polaritons accounting for the spin led to the appearance of \emph{spinoptronics}, an optical analogue of spintronics.\cite{ShelykhSpinopt,PolaritonDevices} It was also shown that spin dependence of polariton-polariton interactions can lead to the appearance of intriguing non-linear polarization phenomena in polariton condensates, such as polarization multistability,\cite{Multistability} full paramagnetic screening also known as spin Meissner effect \cite{SpinMeissner} and spin-dependent condensate velocities in the hybrid Bose-Fermi systems.\cite{Kyriienko} One can expect that the spinor structure of indirect excitons can also have dramatic impact on their collective behavior.\cite{RuboIndEX,ButovSpin}

\section{The model} 
\begin{figure}
\includegraphics[width=1.0\linewidth]{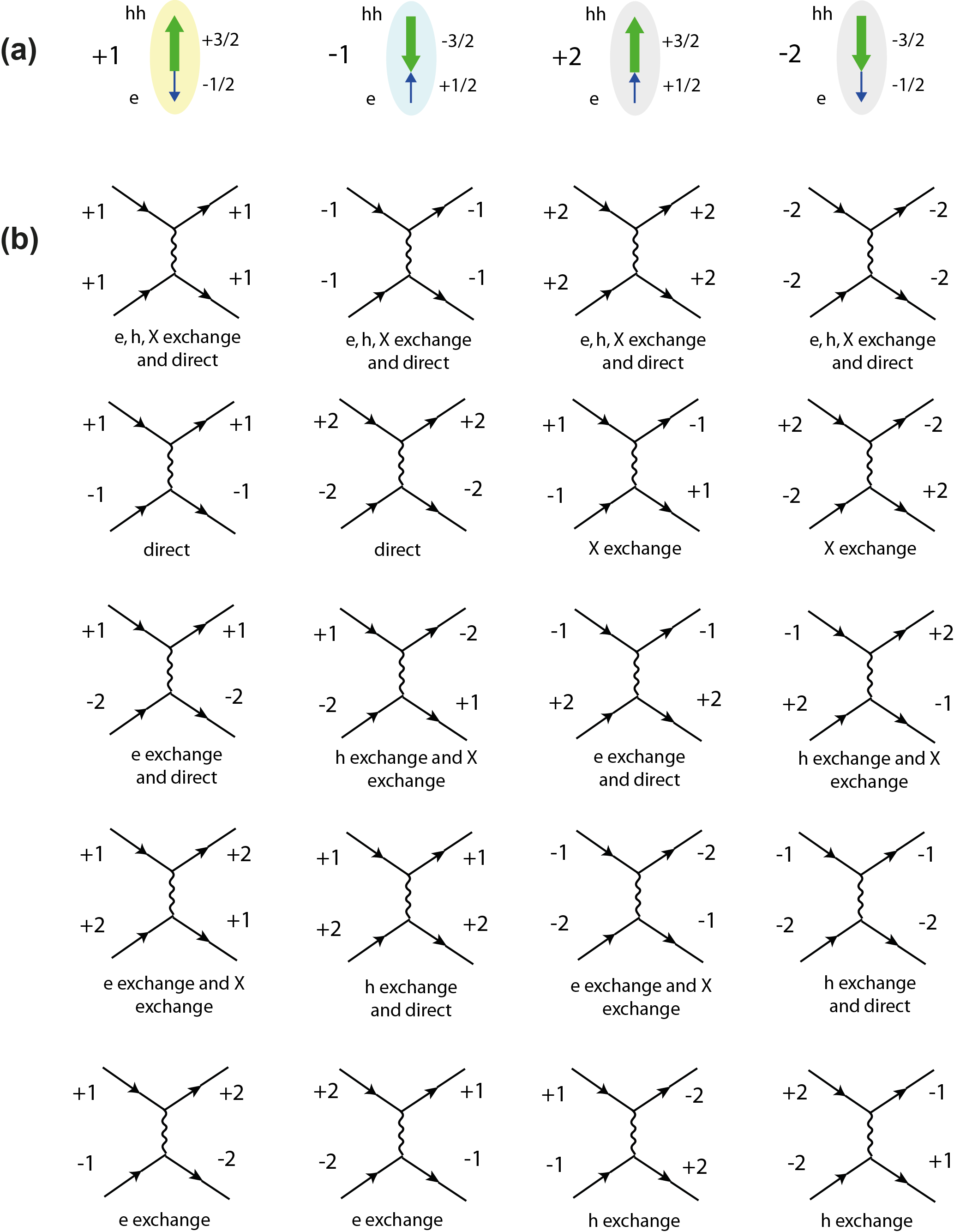}
\caption{(Color online). (a) Spin structure of indirect excitons. (b) Feynmann diagrams showing different types of possible interactions between two excitons.}
\label{Fig1}
\end{figure}

The spin of an indirect exciton is inherited from spins of the individual electron and heavy hole forming it. The possible spin projections of the electron's spin on the structure growth axis (z-axis) are $\pm1/2$, while possible spin projections of heavy holes spin are $\pm3/2$. The exciton thus can have four possible spin projections, $S_z=\pm1,\pm2$ (Fig. \ref{Fig1}, (a)). The  bright states with $S_z=\pm1$ can be created by external right or left circular polarized light, while optical creation of the states with $S_z=\pm2$ is prohibited by selection rules. However, these states known as \emph{dark states} cannot be excluded from the consideration, as they can appear due to the presence of spin-orbit interaction (SOI) of Rashba or Dresselhaus type or can be created as a result of collision of two bright excitons with opposite circular polarizations, as will be discussed below. For direct excitons the energies of bright and dark states are split off by electron-hole exchange interaction with characteristic value of tens of microelectronvolts.\cite{Maialle} However, for indirect excitons, where the overlap between the wave functions of the electrons and holes is very small, this splitting can be neglected and bright and dark excitons can be considered to have the same energy.

Below the temperature of the quantum degeneracy, a system of cold indirect excitons can be thus described by a four component macroscopic wave function  $\Psi(\textbf{r},t)=(\Psi_{+2}(\textbf{r},t),\Psi_{+1}(\textbf{r},t),\Psi_{-1}(\textbf{r},t),\Psi_{-2}(\textbf{r},t))$, where the subscripts correspond to the $z$-projection of the spin. Its dynamics can be obtained from the following equation

\begin{equation}
i\hbar\partial_t\Psi_\sigma=\frac{\delta H}{\delta\Psi^{\ast}_{\sigma}},
\label{EqPsi}
\end{equation}
where $H$ represents the Hamiltonian density of the system accounting for free propagation of particles, SOI and exciton-exciton interactions and can be represented as sum of single-particle and interaction parts $H=H_{0}+H_{int}$. Let us consider the terms $H_{0}$ and $H_{int}$ separately.

The term $H_0$ can be calculated as
\begin{equation}
H_0=\Psi^\dagger(\textbf{r},t)\hat{\textbf{T}}\Psi(\textbf{r},t),
\end{equation}
where the $4\times4$ matrix $\hat{\textbf{T}}$ contains terms corresponding to the kinetic energy and interactions with effective magnetic fields of various types. The latter can be divided into three categories. First, there is SOI acting on the spin of the heavy hole. It scales as a cube of the kinetic momentum \cite{WinklerBook} and we neglect it in further consideration. As for the SOI acting on electrons spin, it can be represented as a sum of the Dresselhaus term $H_D=\beta\left(\sigma_x\hat{k}_x-\sigma_y\hat{k}_y\right)$ arising from the lack of inversion symmetry for the crystalline lattices of most common semiconductor materials (GaAs, CdTe etc) and Rashba term $H_R=\alpha\left(\sigma_x\hat{k}_y-\sigma_y\hat{k}_x\right)$ appearing due to the structural asymmetry of the QW in $z$-direction.\cite{SpintronicsReview} The coefficients $\alpha$ and $\beta$ are constants which depend on the material and geometry of the structure, and $\sigma_{x,y}$ are Pauli matrices. For the indirect excitons both Rashba and Dresselhaus terms lead to the transitions $\pm1\rightarrow\pm2$ mixing bright and dark exciton states and matrix  $\hat{\textbf{T}}$ thus reads
\begin{eqnarray}
\hat{\textbf{T}}=\left(\begin{array}{cc}
  \hat{\textbf{T}}_{12} & 0 \\
  0 & \hat{\textbf{T}}_{12}
\end{array}\right),
\end{eqnarray}
where $2\times2$ blocks $\hat{\textbf{T}}_{12}$ can be written in momentum space representation as
\begin{eqnarray}
\hat{\textbf{T}}^{k}_{12}=\left(\begin{array}{cc}
  \hbar^2 k^2/2M & \hat{S}_{k} \\
  \hat{S}^\ast_{k} & \hbar^2 k^2/2M
\end{array}\right)
\end{eqnarray}
with the operator $\hat{S}_{k}=\chi\left[\beta(\hat{k}_{x}+i\hat{k}_{y})+\alpha(\hat{k}_{y}+i\hat{k}_{x})\right]$.  $\chi=m_{e}/M$ is the ratio of effective electron to exciton masses and mimics the renormalization of indirect exciton dispersions due to spin-orbit interaction of electron (see Appendix A for details). Going to the real space representation one can use the explicit expressions for momentum operators $\hat{k}_{x}=-i\partial_x,\hat{k}_{y}=-i\partial_y$.

Now let us consider the term accounting for exciton-exciton interactions. As indirect exciton is a composite boson,\cite{Combescot,CombescotPRB} they can be divided into four categories, namely the terms corresponding to the direct Coulomb repulsion, exchange of electrons, exchange of holes and simultaneous exchange of electron and hole (exciton exchange). These processes can be visualized using the interaction diagrams shown at Fig. \ref{Fig1} (b).

The corresponding interaction Hamiltonian thus reads
\begin{widetext}
\begin{eqnarray}\nonumber
H_{int}=\frac{V_{dir}+V_{X}+V_{e}+V_{h}}{2}\sum_{\sigma=\pm 1\pm 2}|\Psi_{\sigma}|^4+(V_{dir}+V_{X})(|\Psi_{+1}|^2|\Psi_{-1}|^2+|\Psi_{+2}|^2|\Psi_{-2}|^2)
+(V_{dir}+V_{X}+V_{e}+V_{h})\times \\ \times(|\Psi_{+1}|^2|\Psi_{+2}|^2+|\Psi_{-1}|^2|\Psi_{-2}|^2+|\Psi_{+1}|^2|\Psi_{-2}|^2+|\Psi_{-1}|^2|\Psi_{+2}|^2)+(V_{e}+V_{h})(\Psi^{\ast}_{+1}\Psi^{\ast}_{-1}\Psi_{+2}\Psi_{-2}+\Psi^{\ast}_{+2}\Psi^{\ast}_{-2}\Psi_{+1}\Psi_{-1}),
\label{Hint}
\end{eqnarray}
\end{widetext}
where $V_{dir}$, $V_{X}$, $V_{e}$ and $V_{h}$ denote direct dipole-dipole repulsion, whole exciton exchange, electron exchange and hole exchange Coulomb interaction, respectively. Since we are interested in the behavior of a weakly interacting Bose-Einstein condensate of indirect excitons, the main contribution comes from the processes with zero transferred momentum, and all values of the matrix elements in the above expression are taken for $q=0$. It is well-known that contrary to the case of conventional excitons, the spin-independent direct interaction of indirect excitons does not vanish for zero exchanged momenta due to the strong dipole-dipole repulsion.\cite{Butov1,RuboIndEX,de-Leon} In addition, the processes of electron and hole exchange also have influence on dynamics of the system making it spin-dependent.

The first term of the $H_{int}$ corresponds to the first line of interactions in the Fig. \ref{Fig1} (b) and describes all possible interactions between indirect excitons of the same spin configuration. The second term corresponds to the processes of the direct Coulomb interaction and exciton exchange shown at the second line. The third and fourth lines of interaction diagrams can be combined in the third term of $H_{int}$. Finally, the fifth line in the Fig. \ref{Fig1} (b) corresponds to the fourth term of $H_{int}$ and leads to the transition between pairs of bright and dark indirect excitons.

The interaction constants corresponding to all four types of interaction can be estimated in the same fashion as for direct excitons. The direct Coulomb interaction between indirect exciton is given by
\begin{equation}
V_{dir}=\frac{e^2L}{\epsilon\epsilon_{0}},
\label{V_dir}
\end{equation}
where $L$ is the distance between centres of QWs. The estimation of electron and hole exchange interaction constant is more cumbersome and requires numerical calculation of an exchange integral (\ref{Iexch}) given in the Appendix B (see also Ref. [\onlinecite{de-Leon_2000}]).

\begin{figure}
\includegraphics[width=1.0\linewidth]{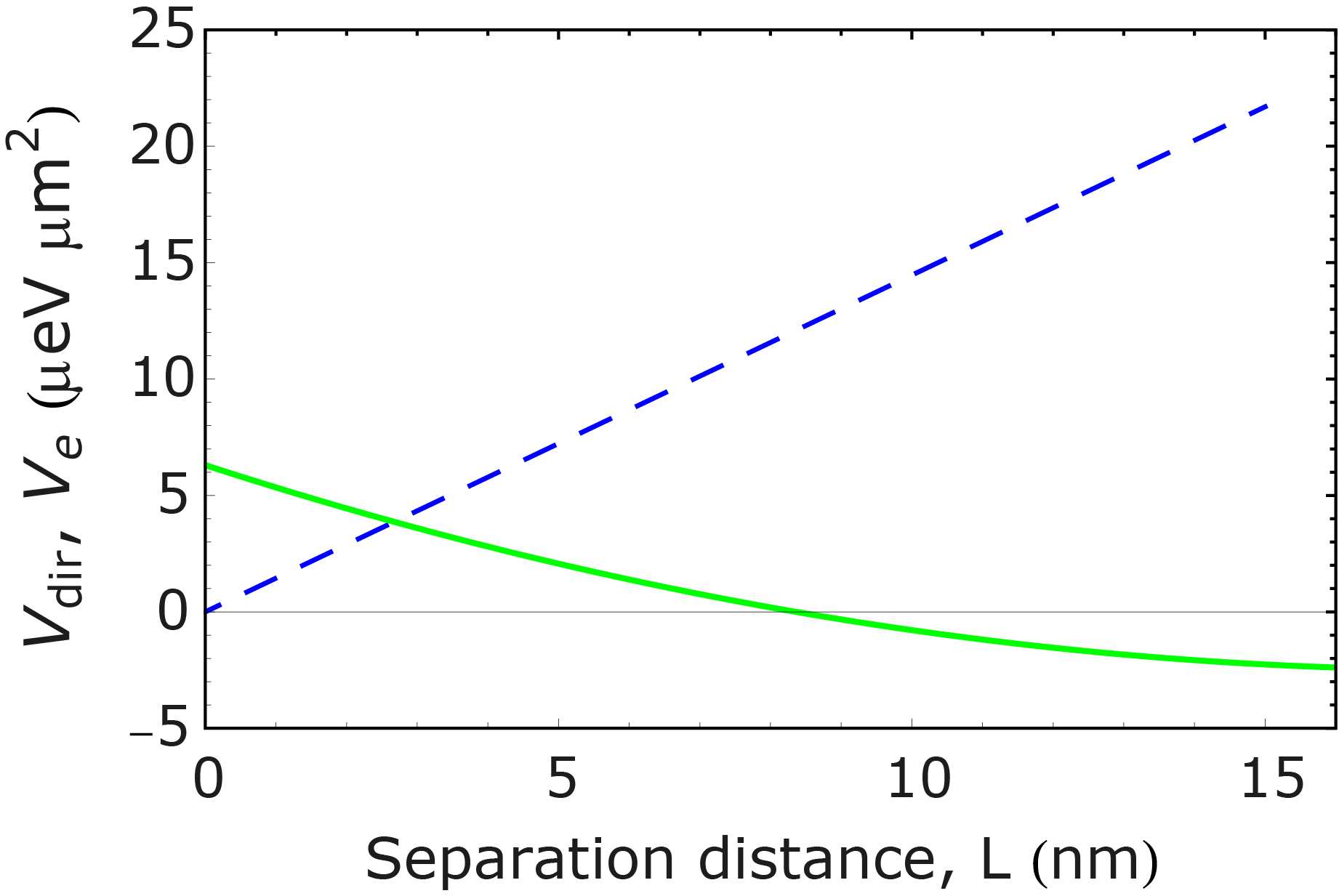}
\caption{(Color online). Electron direct (blue dashed line) and exchange (green solid line) interaction for indirect excitons as a function of separation distance between quantum wells $L$. The sign of the exchange matrix element changes at a certain separation distance. We consider the case of GaAs/AlGaAs/GaAs heterostructure in the narrow QW limit.}
\label{Fig2}
\end{figure}

For indirect excitons in the long wavelength limit ($q\rightarrow 0$) direct and exciton exchange interactions coincide ($V_{dir}=V_{X}$) as well as electron and hole exchange ($V_e=V_h$). It should be noted that the electron and hole exchange interaction strongly depends on the distance between the centres of QWs and changes the sign for certain separation, as it is shown at Fig.\ref{Fig2}). This fact is important for further consideration. 

The set of spinor Gross-Pitaevskii equations describing the dynamics of the system and accounting for both linear terms coming from $H_0$ and non-linear terms provided by particle-particle interactions of various types can be obtained using Eqs. \ref{EqPsi},\ref{Hint} and reads:
\begin{widetext}
\begin{eqnarray}
\label{GPp1}
i\hbar\frac{\partial\Psi_{+1}}{\partial t}=\hat{E}\Psi_{+1}-\hat{S}^{\ast}_{12}\Psi_{+2}+V_{0}\Psi_{+1}|\Psi_{+1}|^{2}+(V_{0}-W)\Psi_{+1}|\Psi_{-1}|^{2}+V_{0}\Psi_{+1}(|\Psi_{-2}|^{2}+|\Psi_{+2}|^{2})+W\Psi^{*}_{-1}\Psi_{+2}\Psi_{-2},\\
\label{GPm1}
i\hbar\frac{\partial\Psi_{-1}}{\partial t}=\hat{E}\Psi_{-1}+\hat{S}_{12}\Psi_{-2}+V_{0}\Psi_{-1}|\Psi_{-1}|^{2}+(V_{0}-W)\Psi_{-1}|\Psi_{+1}|^{2}+V_{0}\Psi_{-1}(|\Psi_{+2}|^{2}+|\Psi_{-2}|^{2})+W\Psi^{*}_{+1}\Psi_{+2}\Psi_{-2},\\
\label{GPp2}
i\hbar\frac{\partial\Psi_{+2}}{\partial t}=\hat{E}\Psi_{+2}+\hat{S}_{12}\Psi_{+1}+V_{0}\Psi_{+2}|\Psi_{+2}|^{2}+(V_{0}-W)\Psi_{+2}|\Psi_{-2}|^{2}+V_{0}\Psi_{+2}(|\Psi_{-1}|^{2}+|\Psi_{+1}|^{2})+W\Psi^{*}_{-2}\Psi_{+1}\Psi_{-1},\\
\label{GPm2}
i\hbar\frac{\partial\Psi_{-2}}{\partial t}=\hat{E}\Psi_{-2}-\hat{S}^{\ast}_{12}\Psi_{-1}+V_{0}\Psi_{-2}|\Psi_{-2}|^{2}+(V_{0}-W)\Psi_{-2}|\Psi_{+2}|^{2}+V_{0}\Psi_{-2}(|\Psi_{+1}|^{2}+|\Psi_{-1}|^{2})+W\Psi^{*}_{+2}\Psi_{+1}\Psi_{-1},
\end{eqnarray}
\end{widetext}
where $V_{0}=V_{dir}+V_{X}+V_{e}+V_{h}$, $W=V_{e}+V_{h}$ and we defined the kinetic energy operator $\hat{E}=-\hbar^{2}\nabla^{2}/2M$. Due to the symmetry between electron and hole exchange at $q=0$ the set of the equations we consider contains only two independent interaction parameters $V_{0},W$. This differs from the case considered in Ref. [\onlinecite{RuboIndEX}] where this symmetry was not accounted for and 5 parameters were used for description of the interactions in the system. In realistic system not all of these parameters will be independent.

\section{Spectrum of elementary excitations}

The calculation of the spectrum of elementary excitations in the system can be done using the linearization of spinor Gross-Pitaevskii equations (\ref{GPp1})-(\ref{GPm2}) with respect to small perturbation around the ground state. The spin configuration of the ground state of the condensate can be found by minimization of its free energy
\begin{equation}
F(\Psi_{+1},\Psi_{-1},\Psi_{+2},\Psi_{-2},\mu)=H-\mu f(\Psi_{+1},\Psi_{-1},\Psi_{+2},\Psi_{-2}),
\label{F}
\end{equation}
where $\mu$ denotes the chemical potential of the condensate and $f(\Psi_{+1},\Psi_{-1},\Psi_{+2},\Psi_{-2})=|\Psi_{+1}|^{2}+|\Psi_{-1}|^{2}+|\Psi_{+2}|^{2}+|\Psi_{-2}|^{2}$. The condition $f(\Psi_{+1},\Psi_{-1},\Psi_{+2},\Psi_{-2})=n$, where $n$ is a total concentration, gives an additional equation for the determination of $\mu$.

Let us investigate the generic Hamiltonian $H=H_{0}+H_{int}$ more precisely. Its first part $H_{0}$ consists of kinetic energy and spin-orbit coupling terms. Both of them depend on the velocity of the particles and can be usually disregarded when considering the ground state, where the interaction Hamiltonian (\ref{Hint}) plays a major role. Nevertheless, we will later show that accounting for SOI leads to qualitative changes of the ground state.

The symmetry of $H_{int}$ with respect to the interactions between components results in a non-trivial ground state solution. Let us consider four particular cases of homogeneous condensates:
\begin{enumerate}[1)]
\item $|\Psi^{0}_{+1}|=\sqrt{n}$, $\Psi^{0}_{-1,\pm 2}=0$ or $|\Psi^{0}_{-1}|=\sqrt{n}$, $\Psi^{0}_{+1,\pm 2}=0$ or $|\Psi^{0}_{+2}|=\sqrt{n}$, $\Psi^{0}_{-2,\pm 1}=0$ or $|\Psi^{0}_{-2}|=\sqrt{n}$, $\Psi^{0}_{+2,\pm 1}=0$ (one-component condensate),
\item $|\Psi^{0}_{+1,-1}|=\sqrt{n/2}$, $\Psi^{0}_{\pm 2}=0$ or $|\Psi^{0}_{+2,-2}|=\sqrt{n/2}$, $\Psi^{0}_{\pm 1}=0$ ("$ii$" two-component condensate),
\item $|\Psi^{0}_{+1,-2}|=\sqrt{n/2}$, $\Psi^{0}_{-1,+2}=0$ or $|\Psi^{0}_{-1,+2}|=\sqrt{n/2}$, $\Psi^{0}_{+1,-2}=0$ ("$ij$" two-component condensate),
\item $|\Psi^{0}_{\pm 1,\pm 2}|=\sqrt{n/4}$ (four-component condensate),
\end{enumerate}
where we defined the ground state wave function $\Psi^{0}_{i}$ for each component. For instance, the first case implies the situation where only one spin component is present in the condensate with total density $n$, while the last case implies equal distribution of the density between the four spin components.

Looking at the interaction Hamiltonian (\ref{Hint}) one can note that the last spin-flip term is important. While all other terms do not depend explicitly on the phase of the condensate wave function, it does. Moreover, it is always possible to make the term $\Psi^{\ast}_{+1}\Psi^{\ast}_{-1}\Psi_{+2}\Psi_{-2}$ negative by adjusting the relative phases of all four condensates, and thus it always lowers the energy of the ground state independently of the sign of electron exchange term $W$. However, the sign of $W$ will still affect the ground state properties since it enters into $V_{0}$ matrix element. Thus, one should consider separately two different situations with negative and positive electron and hole exchange interactions.

For negative $W=V_{e}+V_{h}<0$ (large QWs separations) the free energy of the four condensate states considered above is given by
\begin{align}
\label{Hm1}
&H^{(1)}=(V_{dir}+V_{X}+V_{e}+V_{h})\frac{n^{2}}{2}=\frac{V_{0}n^2}{2}, \\
\label{H2iim}
&H^{(2)}_{ii}=(V_{dir}+V_{X}+\frac{V_{e}+V_{h}}{2})\frac{n^{2}}{2}=(V_{0}-\frac{W}{2})\frac{n^{2}}{2}, \\
\label{H2ijm}
&H^{(2)}_{ij}=(V_{dir}+V_{X}+V_{e}+V_{h})\frac{n^{2}}{2}=\frac{V_{0}n^2}{2}, \\
\label{H4m}
&H^{(4)}=(V_{dir}+V_{X}+V_{e}+V_{h})\frac{n^{2}}{2}=\frac{V_{0}n^2}{2},
\end{align}
where we denoted by $H^{(1)}$, $H^{(2)}_{ii}$, $H^{(2)}_{ij}$ and $H^{(4)}$ the free energy of one-component, "$ii$" two-component, "$ij$" two-component and four-component condensates, respectively.
One can see that the ground state is 7-time degenerate and configurations of one-component, two-component and four-component condensates are possible. The chemical potential in this case is equal to $\mu_{<}=(V_{dir}+V_{X}+V_{e}+V_{h})n=V_{0}n$. Therefore, the ground state of the system will be chosen by spontaneous symmetry breaking mechanism. The high level of the degeneracy of the ground states means that the system can demonstrate a large variety of the topological excitations (solitons, vortices and skyrmions). Moreover, in the system one can in principle observe the fragmentation of the condensate into domains with different spin structure. The analysis of these interesting effects, however, lies beyond the scope of the present paper.

For positive $W=V_{e}+V_{h}>0$ exchange interaction (small QWs separations) the free energy of the system for different types of the condensates  yields
\begin{align}
\label{H1p}
&H^{(1)}=(V_{dir}+V_{X}+V_{e}+V_{h})\frac{n^{2}}{2}=\frac{V_{0}n^2}{2}, \\
\label{H2iip}
&H^{(2)}_{ii}=(V_{dir}+V_{X}+\frac{V_{e}+V_{h}}{2})\frac{n^{2}}{2}=(V_{0}-\frac{W}{2})\frac{n^{2}}{2}, \\
\label{H2ijp}
&H^{(2)}_{ij}=(V_{dir}+V_{X}+V_{e}+V_{h})\frac{n^{2}}{2}=\frac{V_{0}n^2}{2}, \\
\label{H4p}
&H^{(4)}=(V_{dir}+V_{X}+\frac{V_{e}+V_{h}}{2})\frac{n^{2}}{2}=(V_{0}-\frac{W}{2})\frac{n^{2}}{2}.
\end{align}
In this situation the ground state is three times degenerate and ether four-component or two-component condensate in $+1,-1$ or $+2,-2$ configurations is preferable.

The spectrum of elementary excitations in the system can be calculated using the standard method of linearizing the Gross-Pitaevskii equations (\ref{GPp1})-(\ref{GPm2}) with respect to small perturbation taken in form of a plane wave. For instance, in the four component condensate case with equal fraction of each spin state it is taken in the form $\Psi^{0}_{i}=\sqrt{n/4}+A_{i}e^{i(\mathbf{k}\mathbf{r}-\omega t)}+B^{\ast}_{i}e^{-i(\mathbf{k}\mathbf{r}-\omega t)}$.\cite{Pitaevskii} The solution of the system for small amplitudes $A_{i}$ and $B_{i}$ gives the dispersion relations of the quasiparticles in the condensate.

\begin{figure}
\includegraphics[width=0.75\linewidth]{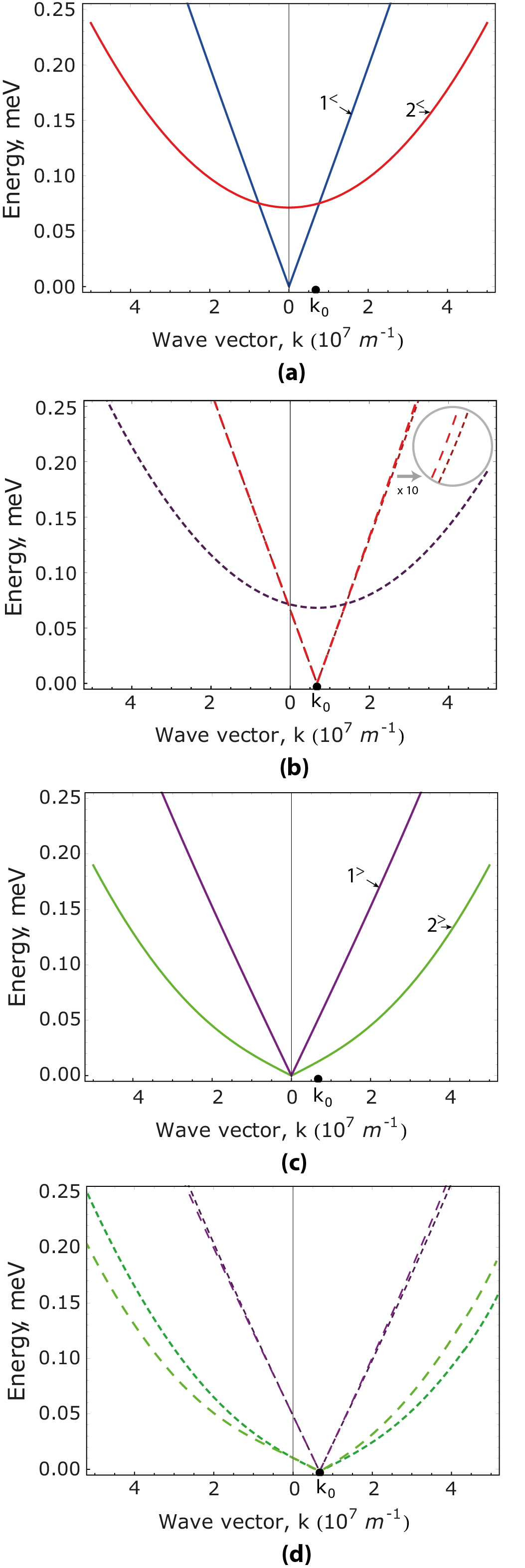}
\caption{(Color online). Quasiparticle modes of a spinor condensate formed by indirect excitons for negative (a, b) and positive (c, d) exchange interaction between excitons. The solid lines in the plots (a) and (c) correspond to the dispersions of the quasiparticles without accounting of the spin-orbit interaction and numbers point out the bare modes described by the formulae (\ref{omega1<})-(\ref{omega2>}). The accounting of SOI removes the spin degeneracy and increases the number of modes. Concentration of the particles is taken as $n=10^{9}$ $cm^{-2}$.}
\label{Fig3}
\end{figure}

First, we consider the case when spin-orbit interaction is absent. For large separation between quantum wells the exchange interaction is attractive and the chemical potential of the ground state is defined as $\mu_{<}=V_{0}n$. The corresponding spectrum of excitations for interaction constants $V_{dir}=19.9$ $\mu eV$ $\mu m^{2}$ and $V_{e}=-1.78$ $\mu eV$ $\mu m^{2}$ is plotted in the Fig. \ref{Fig3}(a). It contains linear Bogoliubov mode, gapped quadratic mode and double degenerated gapless quadratic mode given by relations
\begin{eqnarray}
\label{omega1<}
\hbar\omega_{1}^{<}=\sqrt{E_{k}(E_{k}+2\mu_{<})},\\
\label{omega2<}
\hbar\omega_{2}^{<}=E_{k}+n|W|,\\
\label{omega34<}
\hbar\omega_{3,4}^{<}=E_{k},
\end{eqnarray}
where $E_{k}=\hbar^{2}k^{2}/2M$.

For the case of small separation between QWs the exchange matrix element $V_{e}$ changes sign and becomes repulsive (see Fig. \ref{Fig2}). The chemical potential of the ground state is thus defined as $\mu_{>}=(V_{0}-W/2)n$. We plot the spectrum of elementary excitations for $L=6$ $nm$ separation between wells where $V_{dir}=9.95$ $\mu eV$ $\mu m^{2}$ and $V_{e}=1.24$ $\mu eV$ $\mu m^{2}$ (Fig. \ref{Fig3}(c), solid lines). The dispersion relations have the form
\begin{eqnarray}
\label{omega134>}
\hbar\omega_{1,3,4}^{>}=\sqrt{E_{k}[E_{k}+2\mu_{>}]},\\
\label{omega2>}
\hbar\omega_{2}^{>}=\sqrt{E_{k}(E_{k}+nW)},
\end{eqnarray}
where the modes $\hbar\omega_{1,2,4}$ are now three times degenerate. The obtained spectrum coincides qualitatively with the general dispersion relations given in the Ref. [\onlinecite{RuboIndEX}].

The account of the spin-orbit interaction requires including the Rashba and Dresselhaus terms in the Gross-Pitaevskii equations. One can expect that SOI removes the spin degeneracy and leads to the splitting of the modes. Moreover, due to the spin-orbital interaction one-component condensate ground state is no more preferable since the lowest energy state requires the presence of either $(+1,+2)$ or $(-1,-2)$ components. Here we present the numerical calculation of the quasiparticle spectrum accounting for the isotropic spin-orbit interaction of Rashba type. For non-interacting case the dispersion minimum moves to the $k_{0}=\pm\frac{M\alpha}{\hbar^{2}}$ points which corresponds to the non-zero condensate phase velocity in the ground state (note that group velocity defined as $v_{g}=\hbar^{-1}dE/dk$ remains equal to zero and condensate is not moving). The ground state of the condensate in this case acquires the total non- zero phase $e^{i\textbf{k}_0\textbf{r}}$, where the orientation of $\textbf{k}_0$ vector is defined by spontaneous symmetry breaking process.

The corresponding renormalization of quasiparticle dispersion now occurs in the vicinity of $\textbf{k}_{0}$ points showing the linear spectrum (Fig. \ref{Fig3}(b, d)). This situation is reminiscent to the renormalization of the bogolon dispersions in the exciton-polariton condensate where the role of spin-orbital interaction is played by longitudinal-transverse splitting.\cite{PRLSpinor}

For the negative exchange interaction accounting of Rashba SOI leads to the splitting of degenerate bogolon modes $\hbar\omega_{1,3,4}$, which are linear in the $k=k_{0}$ region and behave as bare SOI modes far from the $k_{0}$ point (Fig. \ref{Fig3}(b), red dashed lines). The gapped mode $\hbar\omega_{2}$ (Fig. \ref{Fig3}(a), red dashed lines) is only slightly renormalized by spin-orbit interaction. One should note that accounting of SOI leads to the ground state formed by "$ij$" two-component condensate or four-component condensate, and rules out the possibility of the formation of single-component condensate as it was mentioned above.

In the case of positive exchange interaction both modes $\hbar\omega_{1,2,4}$ and $\hbar\omega_{2}$ are renormalized in the $k=k_{0}$ point, while for large values of $k$ they approach usual Rashba like dispersion for upper and lower mode $\varepsilon_{\pm}=E_{k}\pm\alpha k$ (Fig. \ref{Fig3}(d)). The ground state in this case corresponds to four-component condensate only.

The similar situation occurs if only Dresselhaus interaction term present in the Hamiltonian. However, the accounting of both Rashba and Dresselhaus SOI requires not trivial ground state definition and we leave it as a subject for future research.

\section{Real space and time dynamics}
The dynamics of the cold exciton droplet can be studied using the set of four Gross-Pitaevskii equations formulated in the previous section (\ref{GPp1})-(\ref{GPm2}). In the present article we do not focus on experiment modelling, but rather give a qualitative examples of the possible different types of the dynamics of concentration and spin of exciton droplets in real space and time. The initial distribution of the excitons in a droplet is modelled by $2$ $\mu m$ diameter Gaussian wavepacket with maximal concentration of the order order $n=10^{9}$ $cm^{-2}$. When thermalized and being far from the hot center, the droplets reveal the physics of a cold boson gas, while an additional indirect exciton supply is provided in the central region by external current through the structure.\cite{ButovKavokin} In the present paper we account for the finite lifetime of the particles ($\tau=2$ $ns$) and study the dynamics of the droplets in stationary and non-stationary regimes.

Let us assume that external optical pump is linear polarized and both $+1$ and $-1$ bright exciton states are created. The kinetic terms in the Gross-Pitaevskii equations for $\Psi_{+1}$ and $\Psi_{-1}$ cause the diffusion of the particles from the center of the spot. The strong dipole-dipole indirect exciton interaction leads to the repulsion of particles from the high concentration regions and manifests itself in formation of concentric propagating rings. While the aforementioned terms are present in the case of exciton polariton condensates and have been widely studied, the additional terms leading to the transitions between bright and dark states are of great importance for the case of the indirect excitons. In the following consideration we describe the effects coming from all terms of this kind separately.

\begin{figure}
\includegraphics[width=1.0\linewidth]{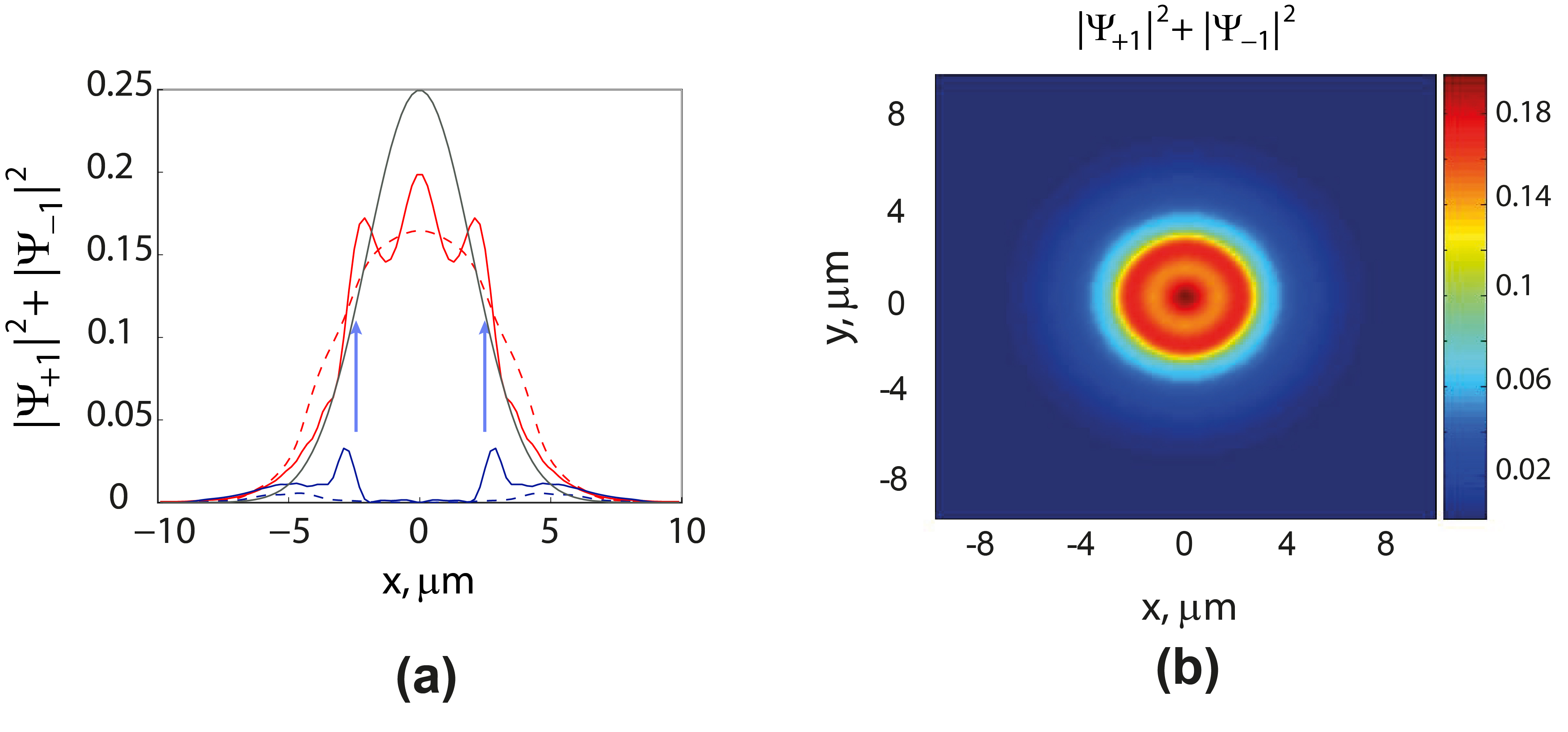}
\caption{(Color online). (a) The bright exciton stationary density profile for the linearly polarized two-component ($+1$ and $-1$) condensate with Dresselhaus SOI present. The spin-orbit interaction constants are taken as $\beta=3$ $\mu eV$ $\mu m$ (red solid line) and $\beta=1.5$ $\mu eV$ $\mu m$ (red dashed line). The grey line represents the initial density profile of the spot (magnitude is reduced by factor of 3), while blue lines point out the regions of the effective bright-to-dark conversion. (b) The 2D density plot for bright excitons showing the "mexican hat" profile ($\beta=3$ $\mu eV$ $\mu m$).}
\label{Fig4}
\end{figure}

Firstly, if only one type of SOI (Rashba or Dresselhaus) is present, for cylindrically symmetric pumping spot the distribution of the intensity of the photoluminescence in the real space governed by the concentration of the bright states remains cylindrically symmetric. The processes of SOI lead to the conversion of the bright states into dark which can be observed in the bright exciton density plots (Fig. \ref{Fig4}).
One can note from the form of SOI operator $\hat{S}_{12}$ that the most efficient conversion of bright states into dark occurs at the points where the density gradient is largest (blue arrows in the Fig. \ref{Fig4}(a)). This yields the "mexican hat" profile in the near field distribution of photoluminescence which depends on SOI strength (solid and dashed lines).

\begin{figure}
\includegraphics[width=1.0\linewidth]{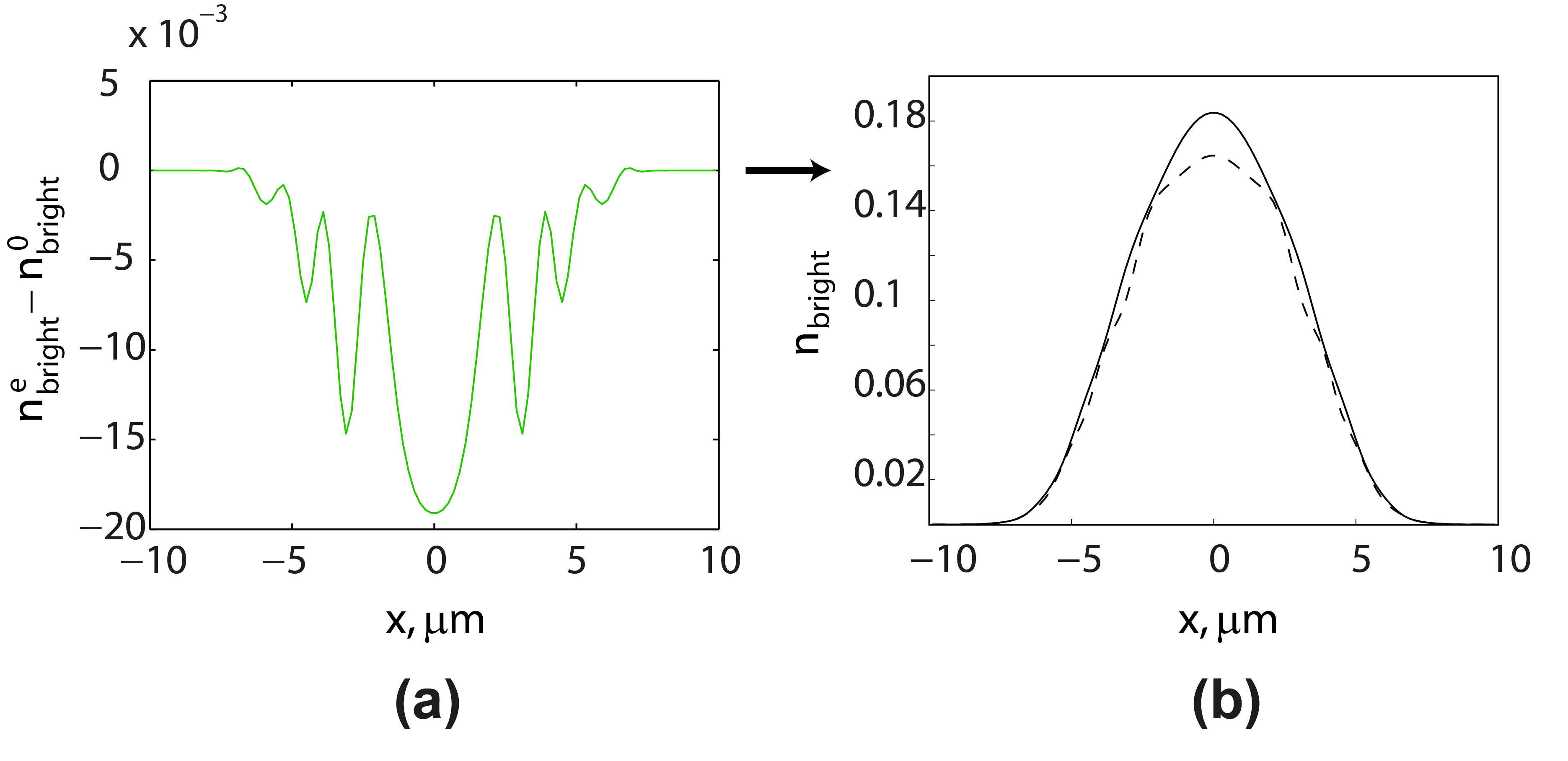}
\caption{(Color online). (a) The effect of bright-dark exciton exchange mixing on real space propagation of cold exciton droplet. The plot shows the difference between distributions of the photoluminescence in cases where  bright-dark exciton exchange is present ($n^{e}_{bright}$) and absent ($n^{0}_{bright}$). (b) The density profile of bright excitons with accounting of exchange bright-to-dark conversion (dashed line) and without accounting of exchange interaction (solid line). The exchange matrix element is equal to $V_{e}=-V_{dir}/5$, which corresponds to the large separation between QWs.}
\label{Fig5}
\end{figure}

Besides SOI, another mechanism can lead to bright to dark exciton conversion. This is the electron or hole exchange interaction term leading to the transitions $\{+1,-1\}\longleftrightarrow\{+2,-2\}$. It is described by the last term in the Gross-Pitaevskii equations (\ref{GPp1})-(\ref{GPm2}) with the interaction constant given by $W=V_{e}+V_{h}$. In the calculations we consider a GaAs/AlGaAs/GaAs structure with $8 nm/4 nm/8 nm$ QWs, where the exchange interaction constant equals to $V_{e}=V_h=-1.3$ $\mu eV$ $\mu m^{2}$. One can see that the most efficient dark exciton creation process takes places for the highest bright exciton density regions. Together with repulsive interaction dynamics, it leads to the appearance of density modulations in the radial direction connected to the rings of bright exciton concentration. The Fig. \ref{Fig5}(a) shows the difference of bright exciton density between the case where the exchange term is present, ($n^{e}_{bright}$),  and case where it is zero, ($n^{0}_{bright}$).
One can see that the exchange term leads to the formation of several dips in the bright exciton density and causes overall flattening of the density profile. For the high density of trapped indirect excitons bright-to-dark conversion could play important role causing the gap in the center of photoluminescence profile.\cite{SnokeTrap} However, in the typically studied structures with indirect excitons where exchange interaction is an order of magnitude smaller than the direct interaction the density modulation is weak and can be expected to be concealed by other factors.

\begin{figure}
\includegraphics[width=1.0\linewidth]{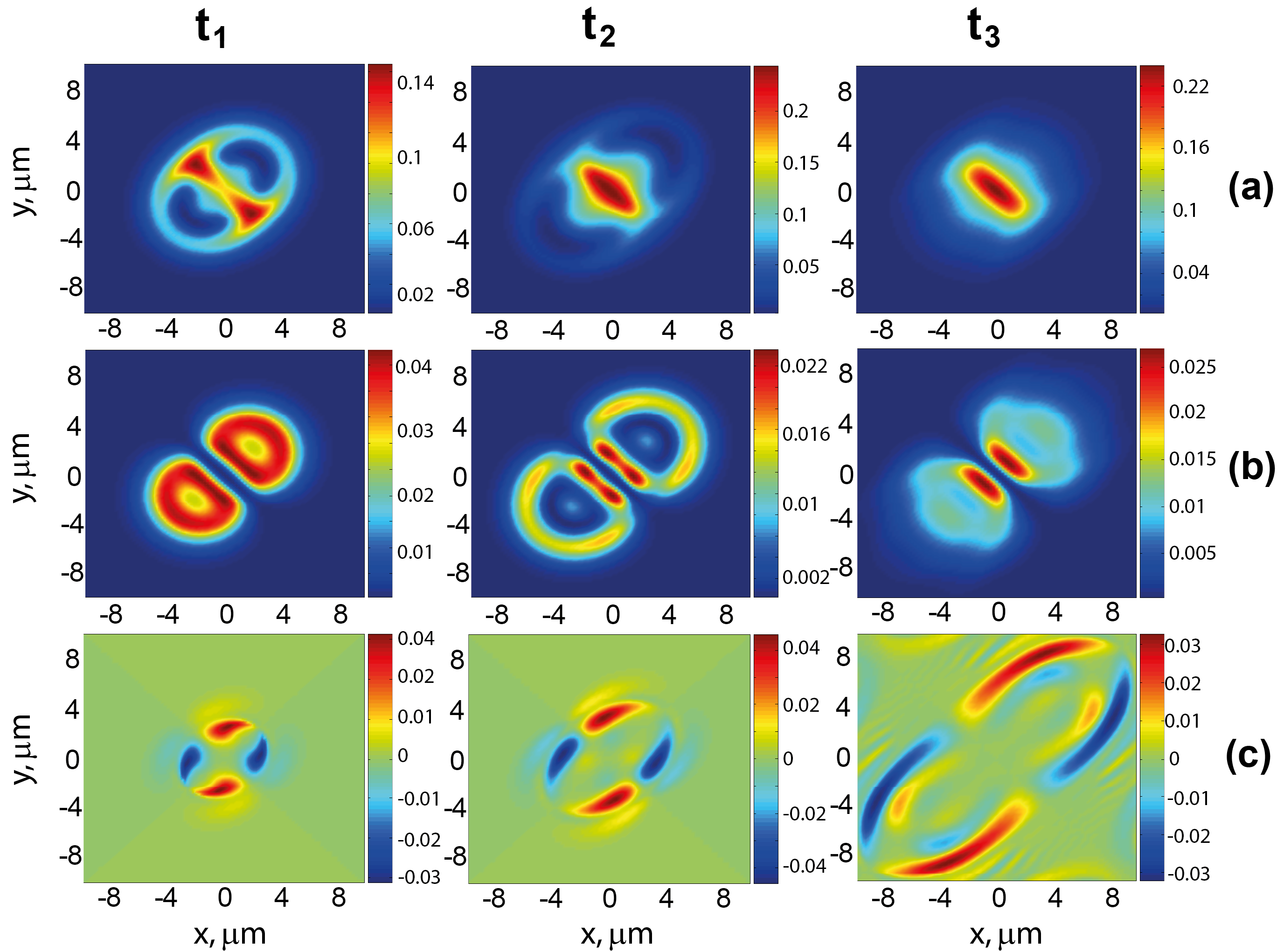}
\caption{(Color online). Dynamics of a cold exciton droplet with both Rashba and Dresselhaus interactions present ($\beta=3$ $\mu eV$ $\mu m$, $\alpha=0.9\beta$). The set of density plots shows the bright exciton (a) and dark exciton (b) density, and the bright circular polarization degree determined as $\wp=(|\Psi_{+1}|^2-|\Psi_{-1}|^2)/(|\Psi_{+1}|^2+|\Psi_{-1}|^2)$ (c).}
\label{Fig6}
\end{figure}

Next, we study the situation when both Rashba and Dresselhaus SOI terms present in the Gross-Pitaevskii equations. in this case the dispersions of non-interacting particles are anisotropic in k-space,
\begin{equation}
\varepsilon_{\pm}(\mathbf{k})=\frac{\hbar^{2}k^{2}}{2m}\pm k\sqrt{\alpha^{2}+\beta^{2}+2\alpha\beta\sin(2\theta_{\mathbf{k}})},
\end{equation}
where $\theta_{\mathbf{k}}$ denotes the angle between the wave vector $\mathbf{k}$ and $x$ axis. This can lead to the breaking of the cylindrical symmetry in the system as it is illustrated in the Fig. \ref{Fig6}.
The density profile in these plots is governed by the interplay of Rashba and Dresselhaus terms, which make it cylindrically non-symmetric. The corresponding evolution of the circular polarization degree shows a four-leaf pattern formation. A similar phenomenon was observed for spots of cold exciton condensates.\cite{ButovKavokin} One can note that in the case of equal strength of Rashba and Dresselhaus SOI the integrated circular polarization is constantly zero.
The spin-orbital interaction strongly depends on the material and geometry of the sample. For the particular $8 nm/4 nm/8 nm$ GaAs/AlGaAs heterostructure the constant of Dresselhaus aSOI can be estimated as $\beta=3$ $\mu eV$ $\mu m$, while Rashba constant can be tuned in wide diapason by the external gate voltage $V_g$.\cite{Sherman,Glazov,SpintronicsReview}

\begin{figure}
\includegraphics[width=1.0\linewidth]{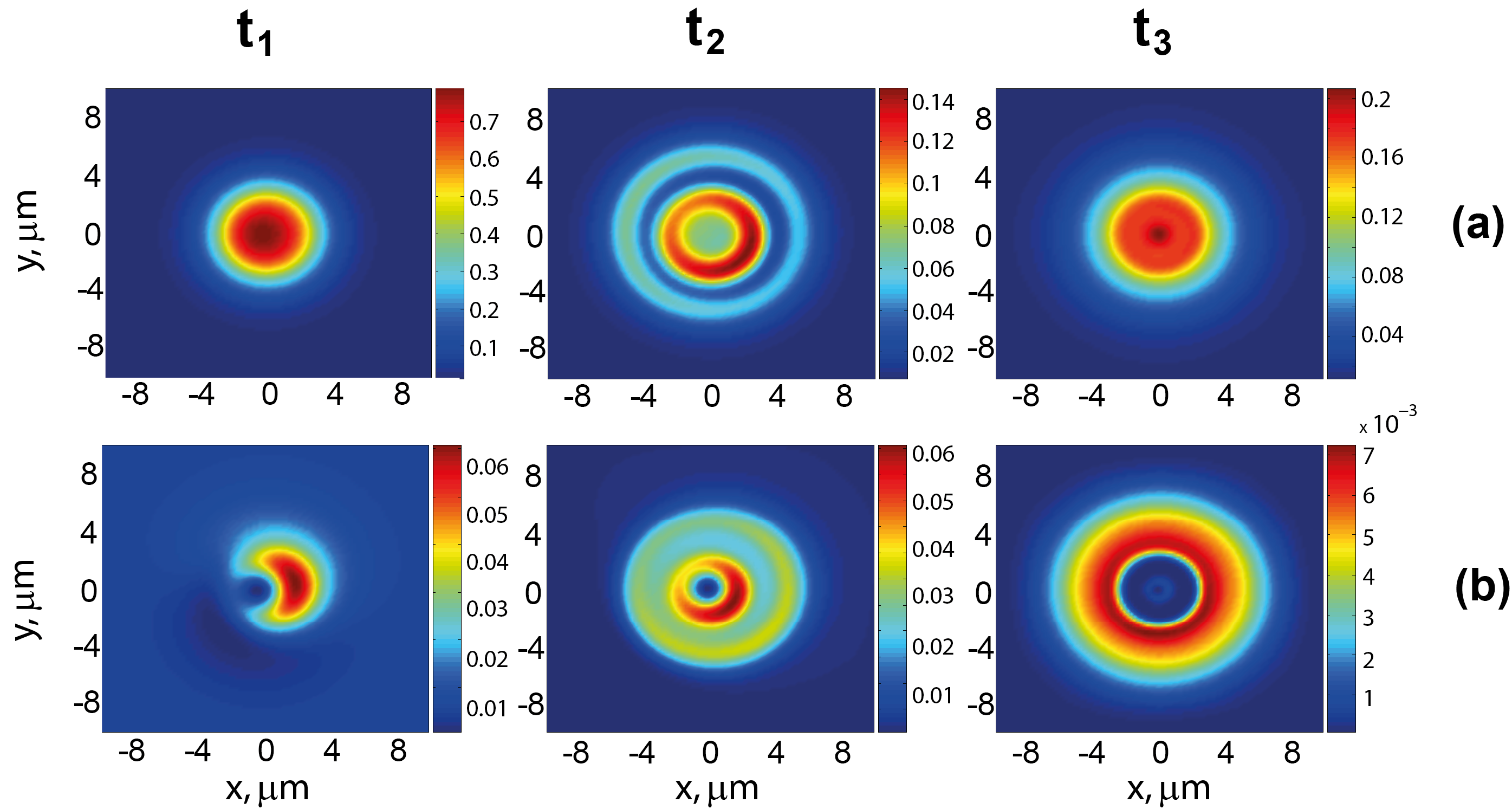}
\caption{(Color online). Dynamics of bright (a) and dark (b) components under the circularly polarized initial conditions. The evolution of exciton density reveals the rotation in time of the velocity field for both species, which can be described as resulting from the appearance of the effective magnetic field provided by the interplay between SOI and .}
\label{Fig7}
\end{figure}

Finally, we consider the case of a circularly polarized condensate where only one component is pumped. In this case the symmetry between $+1\rightarrow+2$ and $-1\rightarrow-2$ conversion is removed and it leads to the appearance of effective magnetic field in z-direction acting on a bright and dark components and arising from spin- dependent exciton-exciton interactions. Its interplay with SOI of Rashba and Dresslhaus types leads to the rotation of the densities of bright and dark components in real space and time as it is shown at Fig. \ref{Fig7}. However, going to the stationary regime the density profile coincides with the situation for a linear pump  (Fig. \ref{Fig7}($t_{3}$)).

\section{Conclusions} In conclusion, we analyzed the ground state properties and spin dynamics in the system of cold indirect excitons, accounting for Rashba and Dresselhaus SOI and spin-dependent exciton-exciton interactions. We demonstrated that ground state of the cold exciton gas is highly degenerate and calculated the dispersions of the Bogoliubov excitations in the system. We have shown that accounting of spin structure qualitatively changes the dynamics of the exciton droplets in the real space and can lead to the formation of the rings and "mexican hat" structures in spatial distribution of near field photoluminescence provided by bright excitons. For the case of both Rashba and Dresselhaus terms present we have shown the four-leave pattern formation for the distribution of circular polarization in the real space.\\

\emph{Acknowledgement.} We thank Yuri G. Rubo and A.V. Kavokin for the valuable discussions. This work was supported by FP7 IRSES projects "SPINMET" and "POLAPHEN" and Rannis "Center of Excellence in Polaritonics". O.K. acknowledges the help from Eimskip Fund. E.B.M. thanks Universidad Autonoma de Mexico for hospitality.

\appendix
\section{Spin-orbit interaction for indirect exciton}

We consider the indirect exciton -- composite boson consisting of electron and hole in separated QWs which are bounded with attractive Coulomb interaction. The generic Hamiltonian of indirect exciton with accounting of SOI for electron (for instance of Rashba type) can be written as
\begin{equation}
\hat{H}=\hat{H}_{0}+V_{int},
\label{H_apnA}
\end{equation}
where $V_{int}=-e^{2}/4\pi\epsilon\epsilon_{0}\sqrt{L^{2}+|\mathbf{r}_{e}-\mathbf{r}_{h}|^{2}}$ denotes the interaction between electron and a hole. The kinetic part of the Hamiltonian now includes the SOI term for an electron and can be represented as $4\times4$ matrix
\begin{eqnarray}
\hat{\textbf{T}}=\left(\begin{array}{cc}
  \hat{\textbf{T}}_{0} & 0 \\
  0 & \hat{\textbf{T}}_{0}
\end{array}\right),
\end{eqnarray}
where $2\times2$ blocks $\hat{\textbf{T}}_{0}$ read
\begin{eqnarray}
\hat{\textbf{T}}_{0}=\left(\begin{array}{cc}
  -\frac{\hbar^{2}}{2m_{e}}\nabla_{e}^{2}-\frac{\hbar^{2}}{2m_{h}}\nabla_{h}^{2} & \alpha(-\partial_{x}^{e}+i\partial_{y}^{e}) \\
  \alpha(\partial_{x}^{e}+i\partial_{y}^{e}) & -\frac{\hbar^{2}}{2m_{e}}\nabla_{e}^{2}-\frac{\hbar^{2}}{2m_{h}}\nabla_{h}^{2}
\end{array}\right)
\label{T_apnA}
\end{eqnarray}
with $m_e$ and $m_h$ being electron and hole mass, respectively, and we denoted the operators acting on electron and a hole by indices $e$ and $h$. The operator $\hat{\textbf{T}}$ acts on the spinor wave function $\Psi=(\Psi_{++},\Psi_{-+},\Psi_{+-},\Psi_{--})$, where indices for each component $\Psi_{ii}$ describe the sign of spin projection on $z$ axis for electron and a hole. Therefore, non-diagonal terms of matrix $\hat{\textbf{T}}_{0}$ couple $\Psi_{++}$ and $\Psi_{-+}$ components and are responsible for the spin flip transitions between bright and dark excitonic states $\pm1\rightarrow\pm2$. For the exciton it results into bright-to-dark states conversion. The operator $\hat{\textbf{T}}_{0}$ can be rewritten in exciton center-of-mass coordinate frame using transformations
\begin{align*}
&\mathbf{R}=\chi \mathbf{r}_{e}+(1-\chi)\mathbf{r}_{h},\\
&\mathbf{r}=\mathbf{r}_{e}-\mathbf{r}_{h},
\end{align*}
where $\chi=m_{e}/(m_{e}+m_{h})$ is the ratio electron and exciton. Thus, the expression (\ref{T_apnA}) reads
\begin{eqnarray}
\hat{\textbf{T}}_{0}=\left(\begin{array}{cc}
  -\frac{\hbar^{2}}{2M}\nabla_{\mathbf{R}}^{2}-\frac{\hbar^{2}}{2\mu}\nabla_{\mathbf{r}}^{2} & \alpha(-\chi\partial_{X}-\partial_{x}+i\chi\partial_{Y}+\partial_{y}) \\
  \alpha(\chi\partial_{X}+\partial_{x}+i\chi\partial_{Y}+\partial_{y}) & -\frac{\hbar^{2}}{2M}\nabla_{\mathbf{R}}^{2}-\frac{\hbar^{2}}{2\mu}\nabla_{\mathbf{r}}^{2}
\end{array}\right),
\label{T_apnA2}
\end{eqnarray}
with $M=m_{e}+m_{h}$ being exciton mass and $\mu=\frac{m_{e}m_{h}}{M}$- reduced mass. Let us for simplicity account only for first matrix $\hat{\textbf{T}}_{0}$ acting on the first pair of states $\Psi_{++}$ and $\Psi_{-+}$. The direct multiplication yields
\begin{widetext}
\begin{align*}
&-\frac{\hbar^{2}}{2M}\nabla_{\mathbf{R}}^{2}\Psi_{++}(\mathbf{R},\mathbf{r})+\chi\alpha(-\partial_{X}+i\partial_{Y})\Psi_{-+}(\mathbf{R},\mathbf{r})+[-\frac{\hbar^{2}}{2\mu}\nabla_{\mathbf{r}}^{2}+V(\mathbf{r})]\Psi_{++}(\mathbf{R},\mathbf{r})+\alpha(-\partial_{x}+i\partial_{y})\Psi_{-+}(\mathbf{R},\mathbf{r})=E\Psi_{++}(\mathbf{R},\mathbf{r}),\\
&-\frac{\hbar^{2}}{2M}\nabla_{\mathbf{R}}^{2}\Psi_{-+}(\mathbf{R},\mathbf{r})+\chi\alpha(\partial_{X}+i\partial_{Y})\Psi_{++}(\mathbf{R},\mathbf{r})+[-\frac{\hbar^{2}}{2\mu}\nabla_{\mathbf{r}}^{2}+V(\mathbf{r})]\Psi_{-+}(\mathbf{R},\mathbf{r})+\alpha(\partial_{x}+i\partial_{y})\Psi_{++}(\mathbf{R},\mathbf{r})=E\Psi_{-+}(\mathbf{R},\mathbf{r}).
\end{align*}
\end{widetext}
Substituting the wave functions in the plane wave form $\Psi_{++}(\mathbf{R},\mathbf{r})=e^{i\mathbf{K}\mathbf{R}}\phi_{++}(\mathbf{r})$ and $\Psi_{-+}(\mathbf{R},\mathbf{r})=e^{i\mathbf{K}\mathbf{R}}\phi_{-+}(\mathbf{r})$ one can see that the Rashba SOI affects the center-of-mass motion of the indirect exciton. Thereby, the accounting of spin-orbit interaction for electron leads to the appearance of SOI acting on the indirect exciton reduced by a factor $\chi=m_{e}/M$ denoting the electron-exciton mass ratio. The influence of SOI on the relative electron-hole motion can be studied within the perturbation theory and one can show that for the $1s$ state of exciton the first order correction is zero. Therefore, in the first order approximation one can neglect relative motion terms.

\section{Interaction matrix elements for indirect excitons}

The problem of calculation of indirect exciton interaction constant is reminiscent to the same problem for direct excitons. However, one should account that in the former case electron and hole layers are separated by certain distance and the wave functions of indirect exciton are modified. The indirect exciton interaction matrix elements were calculated in Ref. [\onlinecite{de-Leon}]. However, in the following calculations we will follow the approach used in Ref. [\onlinecite{Ciuti}] for direct excitons to write integrals in the similar form. As was stated before, the spin-dependent exciton interactions are based on four types of Feynman diagrams: direct interaction, exciton exchange, electron exchange and hole exchange. These interactions can be written in general form
\begin{widetext}
\begin{align}
\label{Vdir}
V_{dir}(\mathbf{Q},\mathbf{Q'},\mathbf{q})&=& \int d^{2}\mathbf{r_{e}} d^{2}\mathbf{r_{h}} d^{2}\mathbf{r_{e'}} d^{2}\mathbf{r_{h'}} \Psi^{\ast}_{\mathbf{Q}}(\mathbf{r_{e}},\mathbf{r_{h}})\Psi^{\ast}_{\mathbf{Q'}}(\mathbf{r_{e}},\mathbf{r_{h'}})V_{I}(\mathbf{r_{e}},\mathbf{r_{h}},\mathbf{r_{e'}},\mathbf{r_{h'}})\Psi_{\mathbf{Q+q}}(\mathbf{r_{e}},\mathbf{r_{h}})\Psi_{\mathbf{Q'-q}}(\mathbf{r_{e'}},\mathbf{r_{h'}}),\\
\label{Vx}
V^{X}_{exch}(\mathbf{Q},\mathbf{Q'},\mathbf{q})&=& \int d^{2}\mathbf{r_{e}} d^{2}\mathbf{r_{h}} d^{2}\mathbf{r_{e'}} d^{2}\mathbf{r_{h'}} \Psi^{\ast}_{\mathbf{Q}}(\mathbf{r_{e}},\mathbf{r_{h}})\Psi^{\ast}_{\mathbf{Q'}}(\mathbf{r_{e}},\mathbf{r_{h'}})V_{I}(\mathbf{r_{e}},\mathbf{r_{h}},\mathbf{r_{e'}},\mathbf{r_{h'}})\Psi_{\mathbf{Q+q}}(\mathbf{r_{e'}},\mathbf{r_{h'}})\Psi_{\mathbf{Q'-q}}(\mathbf{r_{e}},\mathbf{r_{h}}),\\
\label{Vexch_e}
V^{e}_{exch}(\mathbf{Q},\mathbf{Q'},\mathbf{q})&=& -\int d^{2}\mathbf{r_{e}} d^{2}\mathbf{r_{h}} d^{2}\mathbf{r_{e'}} d^{2}\mathbf{r_{h'}} \Psi^{\ast}_{\mathbf{Q}}(\mathbf{r_{e}},\mathbf{r_{h}})\Psi^{\ast}_{\mathbf{Q'}}(\mathbf{r_{e}},\mathbf{r_{h'}})V_{I}(\mathbf{r_{e}},\mathbf{r_{h}},\mathbf{r_{e'}},\mathbf{r_{h'}})\Psi_{\mathbf{Q+q}}(\mathbf{r_{e'}},\mathbf{r_{h}})\Psi_{\mathbf{Q'-q}}(\mathbf{r_{e}},\mathbf{r_{h'}}),\\
\label{Vexch_h}
V^{h}_{exch}(\mathbf{Q},\mathbf{Q'},\mathbf{q})&=& -\int d^{2}\mathbf{r_{e}} d^{2}\mathbf{r_{h}} d^{2}\mathbf{r_{e'}} d^{2}\mathbf{r_{h'}} \Psi^{\ast}_{\mathbf{Q}}(\mathbf{r_{e}},\mathbf{r_{h}})\Psi^{\ast}_{\mathbf{Q'}}(\mathbf{r_{e}},\mathbf{r_{h'}})V_{I}(\mathbf{r_{e}},\mathbf{r_{h}},\mathbf{r_{e'}},\mathbf{r_{h'}})\Psi_{\mathbf{Q+q}}(\mathbf{r_{e}},\mathbf{r_{h'}})\Psi_{\mathbf{Q'-q}}(\mathbf{r_{e'}},\mathbf{r_{h}}),
\end{align}
\end{widetext}
where the Coulomb interaction between the electrons and holes of different excitons are
\begin{align*}
V_{I}(\mathbf{r_{e}},\mathbf{r_{h}},\mathbf{r_{e'}},\mathbf{r_{h'}})&=V(|\mathbf{r_{e}}-\mathbf{r_{e'}}|)+V(|\mathbf{r_{h}}-\mathbf{r_{h'}}|)-\\&-V(|\mathbf{r_{e}}-\mathbf{r_{h'}}|)-V(|\mathbf{r_{e'}}-\mathbf{r_{h}}|),
\label{V_I}
\end{align*}
where minus sign for last two terms account for attractive interaction between an electron and a hole.

The Coulomb interaction for indirect excitons can be written as
\begin{align*}
V_{I}(\mathbf{r_{e}},\mathbf{r_{h}},\mathbf{r_{e'}},\mathbf{r_{h'}})=\frac{e^2}{4\pi\epsilon\epsilon_{0}}\Big[\frac{1}{|\mathbf{r_{e}}-\mathbf{r_{e'}}|}+\frac{1}{|\mathbf{r_{h}}-\mathbf{r_{h'}}|}-\\-\frac{1}{\sqrt{(\mathbf{r_{e}}-\mathbf{r_{h'}})^{2}+L^{2}}}-\frac{1}{\sqrt{(\mathbf{r_{h}}-\mathbf{r_{e'}})^{2}+L^{2}}}\Big],
\end{align*}
where $L$ is a separation distance between centers of coupled QWs and we used narrow QW approximation.

There are several possibilities to construct the wave function of indirect exciton. As in the usual case of direct excitons we consider the 2D motion of bounded electron-hole pair but with fixed separation in $z$ direction equal to $L$. In this case it is convenient to separate the exciton center of mass motion and relative motion. The general form of wave function is
\begin{equation*}
f(\mathbf{r}_{\parallel},\mathbf{R}_{\parallel})=\frac{e^{i\mathbf{K}_{\parallel}\mathbf{R}_{\parallel}}}{\sqrt{A}}\phi(\mathbf{r}_{\parallel}),
\end{equation*}
where new coordinates are $\mathbf{r}_{\parallel}=\mathbf{r}^{e}_{\parallel}-\mathbf{r}^{h}_{\parallel}$ describing relative motion and $\mathbf{R}_{\parallel}=\beta_{e}\mathbf{r}^{e}_{\parallel}+\beta_{h}\mathbf{r}^{h}_{\parallel}$ for exciton center-of-mass motion. Here $\beta_{e}=m_{e}/(m_{e}+m_{h})$, $\beta_{h}=m_{h}/(m_{e}+m_{h})$ and $A$ denotes the area of the sample.
While the center of mass motion is described by the plane wave function, the relative motion part of wave function $\phi(\mathbf{r}_{\parallel})$ can be represented in several different forms\cite{de-Leon_2000}
\begin{align}
\label{phi1}
&\phi_{1}(\mathbf{r}_{\parallel})=\sqrt{\frac{2}{\pi}}\frac{1}{a_B}\exp\Big(-\frac{|\mathbf{r}_{\parallel}|}{a_B}\Big),\\
\label{phi2}
&\phi_{2}(\mathbf{r}_{\parallel})=\frac{1}{\sqrt{2\pi b(b+r_{0})}}\exp\Big(-\frac{\sqrt{r^{2}_{\parallel}+r_{0}^{2}}-r_{0}}{2b}\Big),
\end{align}
where $a_B$ and $2b$ are quantities associated to indirect exciton Bohr radii obtained by the variational procedure and $r_{0}$ is variational parameter reminiscent to the separation distance between QWs.

\emph{Direct and exciton exchange interaction of indirect excitons.}
To calculate the direct dipole-dipole interaction of two indirect excitons we choose the second representation of the wave function with $\phi(\mathbf{r}_{\parallel})=\phi_{2}(\mathbf{r}_{\parallel})$. In the following derivation we will omit sign $\parallel$ meaning everywhere the 2D motion. Thus, the integral can be written as
\begin{widetext}
\begin{align}
\notag
&V_{dir}(\mathbf{Q},\mathbf{Q'},\mathbf{q})=\frac{e^{2}}{4\pi\epsilon\epsilon_{0}A^{2}}\frac{\exp(2r_{0}/b)}{(2\pi b(b+r_{0}))^{2}}\int d^{2}\mathbf{r}_{e}d^{2}\mathbf{r}_{h}d^{2}\mathbf{r}_{e'}d^{2}\mathbf{r}_{h'} \exp\Big(-\frac{\sqrt{(\mathbf{r}_{e}-\mathbf{r}_{h})^2+r_{0}^{2}}}{b}\Big) \exp\Big(-\frac{\sqrt{(\mathbf{r}_{e'}-\mathbf{r}_{h'})^2+r_{0}^{2}}}{b}\Big) \\ &\exp[-i\mathbf{Q}(\beta_{e}\mathbf{r}_{e}+\beta_{h}\mathbf{r}_{h})]\exp[-i\mathbf{Q'}(\beta_{e}\mathbf{r}_{e'}+\beta_{h}\mathbf{r}_{h'})] \Big[\frac{1}{|\mathbf{r}_{e}-\mathbf{r}_{e'}|}+\frac{1}{|\mathbf{r}_{h}-\mathbf{r}_{h'}|}-\frac{1}{\sqrt{(\mathbf{r_{e}}-\mathbf{r_{h'}})^{2}+L^{2}}}-\frac{1}{\sqrt{(\mathbf{r_{h}}-\mathbf{r_{e'}})^{2}+L^{2}}}\Big]\times \notag \\ &\times\exp[i(\mathbf{Q}+\mathbf{q})(\beta_{e}\mathbf{r}_{e}+\beta_{h}\mathbf{r}_{h})]\exp[i(\mathbf{Q'}-\mathbf{q})(\beta_{e}\mathbf{r}_{e'}+\beta_{h}\mathbf{r}_{h'})].
\label{Vdir_1}
\end{align}
\end{widetext}
This integral can be simplified if one introduces center-of-motion coordinates for both indirect excitons: $\mathbf{R}=\beta_{e}\mathbf{r}_{e}+\beta_{h}\mathbf{r}_{h}$, $\mathbf{R'}=\beta_{e}\mathbf{r}_{e'}+\beta_{h}\mathbf{r}_{h'}$, $\mathbf{\boldsymbol\rho}=\mathbf{r}_{e}-\mathbf{r}_{h}$ and $\mathbf{\boldsymbol\rho'}=\mathbf{r}_{e'}-\mathbf{r}_{h'}$. Then Eq. (\ref{Vdir_1}) yields
\begin{widetext}
\begin{align}\notag
&V_{dir}(\mathbf{q})=\frac{e^{2}}{4\pi\epsilon\epsilon_{0}A^{2}}\frac{\exp(2r_{0}/b)}{(2\pi b(b+r_{0}))^{2}}\int d^{2}\mathbf{\boldsymbol\rho}d^{2}\mathbf{\boldsymbol\rho'}d^{2}\mathbf{R}d^{2}\mathbf{R'} \exp\Big(-\frac{\sqrt{\mathbf{\boldsymbol\rho}^{2}+r_{0}^{2}}}{b}\Big) \exp\Big(-\frac{\sqrt{\mathbf{\boldsymbol\rho'}^{2}+r_{0}^{2}}}{b}\Big)\exp[i\mathbf{q}(\mathbf{R}-\mathbf{R'})]\times \\ \notag  &\times\Big[\frac{1}{|\beta_{h}(\mathbf{\boldsymbol\rho}-\mathbf{\boldsymbol\rho'})+\mathbf{R}-\mathbf{R'}|}+\frac{1}{|-\beta_{e}(\mathbf{\boldsymbol\rho}-\mathbf{\boldsymbol\rho'})+\mathbf{R}-\mathbf{R'}|}-\frac{1}{\sqrt{(\beta_{h}\mathbf{\boldsymbol\rho}+\beta_{e}\mathbf{\boldsymbol\rho'}+\mathbf{R}-\mathbf{R'})^{2}+L^{2}}}- \\ &-\frac{1}{\sqrt{(-\beta_{e}\mathbf{\boldsymbol\rho}-\beta_{h}\mathbf{\boldsymbol\rho'}+\mathbf{R}-\mathbf{R'})^{2}+L^{2}}}\Big],
\label{Vdir_2}
\end{align}
\end{widetext}
where one can note that complex exponents with $\mathbf{Q}$ and $\mathbf{Q'}$ cancel each other. It is convenient to use the next substitutions $\boldsymbol\xi=\mathbf{R}-\mathbf{R'}$, $\boldsymbol\sigma=(\mathbf{R}+\mathbf{R'})/2$. The integral (\ref{Vdir_2}) rewritten in a new variables reads
\begin{widetext}
\begin{align}
\notag
V_{dir}(\mathbf{q})&=\frac{e^{2}}{4\pi\epsilon\epsilon_{0}A}\frac{\exp(2r_{0}/b)}{(2\pi b(b+r_{0}))^{2}}\int d^{2}\mathbf{\boldsymbol\rho}d^{2}\mathbf{\boldsymbol\rho'}d^{2}\mathbf{\boldsymbol\xi} \exp\Big(-\frac{\sqrt{\mathbf{\boldsymbol\rho}^{2}+r_{0}^{2}}}{b}\Big) \exp\Big(-\frac{\sqrt{\mathbf{\boldsymbol\rho'}^{2}+r_{0}^{2}}}{b}\Big)\exp[i\mathbf{q}\boldsymbol\xi]\times \\ &\times \Big[\frac{1}{|\beta_{h}(\mathbf{\boldsymbol\rho}-\mathbf{\boldsymbol\rho'})+\boldsymbol\xi|}+\frac{1}{|-\beta_{e}(\mathbf{\boldsymbol\rho}-\mathbf{\boldsymbol\rho'})+\boldsymbol\xi|}-\frac{1}{\sqrt{(\beta_{h}\mathbf{\boldsymbol\rho}+\beta_{e}\mathbf{\boldsymbol\rho'}+\boldsymbol\xi)^{2}+L^{2}}}-\frac{1}{\sqrt{(-\beta_{e}\mathbf{\boldsymbol\rho}-\beta_{h}\mathbf{\boldsymbol\rho'}+\boldsymbol\xi)^{2}+L^{2}}}\Big],
\label{Vdir_3}
\end{align}
\end{widetext}
where the two dimensional integral over variable $\boldsymbol\sigma$ gives the area $A$. The expression (\ref{Vdir_3}) represents a sum of four integrals for electron-electron, hole-hole and electron-hole mixed interaction
\begin{equation}
V_{dir}=C\Big[\mathcal{I}_{ee'}+\mathcal{I}_{hh'}-\mathcal{I}_{eh'}-\mathcal{I}_{he'} \Big]
\label{Vdir_4}
\end{equation}
where we defined the constant $C=\frac{e^{2}}{4\pi\epsilon\epsilon_{0}A}\frac{\exp(2r_{0}/b)}{(2\pi b(b+r_{0}))^{2}}$.

Let us calculate all integrals separately. The first integral $\mathcal{I}_{ee'}$ yields
\begin{widetext}
\begin{align}
\notag
\mathcal{I}_{ee'}(\mathbf{q})&=C\int d^{2}\mathbf{\boldsymbol\rho}d^{2}\mathbf{\boldsymbol\rho'}d^{2}\mathbf{\boldsymbol\xi}
\exp\Big(-\frac{\sqrt{\mathbf{\boldsymbol\rho}^{2}+r_{0}^{2}}}{b}\Big) \exp\Big(-\frac{\sqrt{\mathbf{\boldsymbol\rho'}^{2}+r_{0}^{2}}}{b}\Big)\exp[i\mathbf{q}\boldsymbol\xi]\Big[\frac{1}{\sqrt{\beta^{2}_{h}(\mathbf{\boldsymbol\rho}-\mathbf{\boldsymbol\rho'})^{2}+\boldsymbol\xi^{2}}}\Big]= \\ \notag &=C\int d^{2}\mathbf{\boldsymbol\rho}d^{2}\mathbf{\boldsymbol\rho'}\exp\Big(-\frac{\sqrt{\mathbf{\boldsymbol\rho}^{2}+r_{0}^{2}}}{b}\Big) \exp\Big(-\frac{\sqrt{\mathbf{\boldsymbol\rho'}^{2}+r_{0}^{2}}}{b}\Big)\int_{0}^{+\infty}\frac{\xi d\xi}{\sqrt{a^2+\xi^{2}}}\int_{0}^{2\pi}d\phi \exp(iq\xi \cos\phi)= \\ \label{Ie1e2} &=2\pi C\int d^{2}\mathbf{\boldsymbol\rho}d^{2}\mathbf{\boldsymbol\rho'} \exp\Big(-\frac{\sqrt{\mathbf{\boldsymbol\rho}^{2}+r_{0}^{2}}}{b}\Big) \exp\Big(-\frac{\sqrt{\mathbf{\boldsymbol\rho'}^{2}+r_{0}^{2}}}{b}\Big)\int_{0}^{+\infty}d\xi\frac{\xi J_{0}(q\xi)}{\sqrt{a^2+\xi^2}}=\\ &=\frac{2\pi}{q}C\int d^{2}\mathbf{\boldsymbol\rho}d^{2}\mathbf{\boldsymbol\rho'} \exp\Big(-\frac{\sqrt{\mathbf{\boldsymbol\rho}^{2}+r_{0}^{2}}}{b}\Big) \exp\Big(-\frac{\sqrt{\mathbf{\boldsymbol\rho'}^{2}+r_{0}^{2}}}{b}\Big) \exp[-\beta_{h}\mathbf{q}\boldsymbol\rho]\exp[\beta_{h}\mathbf{q}\boldsymbol\rho'],
\notag
\end{align}
\end{widetext}
where we made substitution $a^{2}=\beta^{2}_{h}(\mathbf{\boldsymbol\rho}-\mathbf{\boldsymbol\rho'})^{2}$.
The integrals on $\boldsymbol\rho$ and $\boldsymbol\rho'$ are identical and can be factorized into $\mathcal{I}_{ee'}(\mathbf{q})=\mathcal{I}_{\rho}^{2}(\mathbf{q})$, where
\begin{align}
\notag
\mathcal{I}_{\rho}(\mathbf{q})&=\int d\rho\rho\exp\Big(-\frac{\sqrt{\rho^{2}+r_{0}^{2}}}{b}\Big)\int_{0}^{2\pi}e^{-\beta_{h}q\rho}=\\
&=2\pi\int d\rho\rho\exp\Big(-\frac{\sqrt{\rho^{2}+r_{0}^{2}}}{b}\Big)J_{0}(\beta_{h}q\rho).
\label{Irho}
\end{align}
It is not possible to calculate integral (\ref{Irho}) analytically in general case, but we are interested in $q=0$ limit. Then, the Bessel function of zero order $J_{0}=1$ and we can change the variable to $x^{2}=\rho^{2}+r_{0}^2$ and integrate by parts
\begin{align*}
\mathcal{I}_{\rho}(\mathbf{q}\rightarrow 0)&=2\pi\int_{0}^{+\infty} d\rho\rho\exp\Big(-\frac{\sqrt{\rho^{2}+r_{0}^{2}}}{b}\Big)=\\ &=2\pi\int_{r_{0}}^{+\infty} dx\cdot x e^{-x/b}=2\pi e^{-r_{0}/b}b(b+r_{0}).
\end{align*}
Finally, the integral $\mathcal{I}_{ee'}$ reads
\begin{align*}
\mathcal{I}_{ee'}^{\mathbf{q}\rightarrow 0}&=\frac{e^{2}}{4\pi\epsilon\epsilon_{0}A}\frac{\exp(2r_{0}/b)}{(2\pi b(b+r_{0}))^{2}}\frac{2\pi}{q}(2\pi e^{-r_{0}/b}b(b+r_{0}))^{2}=\\ &=\frac{e^2}{2\pi\epsilon\epsilon_{0}Aq}.
\end{align*}

One can see that expression for integral $\mathcal{I}_{hh'}$ coincides with $\mathcal{I}_{ee'}$ with substitution $\beta_{h}\rightarrow \beta_{e}$. Therefore, in the $q\rightarrow 0$ limit they are equal $\mathcal{I}_{hh'}=e^2/2\pi\epsilon\epsilon_{0}Aq$.

Now let us calculate the second type of integrals responsible for the electron-hole attractive interaction
\begin{align*}
&\mathcal{I}_{eh'}(\mathbf{q})=C\int d^{2}\mathbf{\boldsymbol\rho}d^{2}\mathbf{\boldsymbol\rho'}d^{2}\mathbf{\boldsymbol\xi}
\exp\Big(-\frac{\sqrt{\mathbf{\boldsymbol\rho}^{2}+r_{0}^{2}}}{b}\Big)\times \\ &\times \exp\Big(-\frac{\sqrt{\mathbf{\boldsymbol\rho'}^{2}+r_{0}^{2}}}{b}\Big)\exp[i\mathbf{q}\boldsymbol\xi]\Big[\frac{1}{\sqrt{(\beta_{h}\mathbf{\boldsymbol\rho}+\beta_{e}\mathbf{\boldsymbol\rho'}+\boldsymbol\xi)^{2}+L^{2}}}\Big],
\end{align*}
and performing the substitution $\boldsymbol\chi=\beta_{h}\mathbf{\boldsymbol\rho}+\beta_{e}\mathbf{\boldsymbol\rho'}+\boldsymbol\xi$ we can write
\begin{widetext}
\begin{align}\notag
&\mathcal{I}_{eh'}(\mathbf{q})=C\int d^{2}\mathbf{\boldsymbol\rho}d^{2}\mathbf{\boldsymbol\rho'}d^{2}\mathbf{\boldsymbol\chi}
\exp\Big(-\frac{\sqrt{\mathbf{\boldsymbol\rho}^{2}+r_{0}^{2}}}{b}\Big) \exp\Big(-\frac{\sqrt{\mathbf{\boldsymbol\rho'}^{2}+r_{0}^{2}}}{b}\Big)\exp[-i\mathbf{q}\boldsymbol\chi]\exp[-i\beta_{h}\mathbf{q}\boldsymbol\rho]
\exp[-i\beta_{e}\mathbf{q}\boldsymbol\rho'] \Big[\frac{1}{\sqrt{\boldsymbol\chi^{2}+L^{2}}}\Big]= \\ \notag
&=C\int d^{2}\mathbf{\boldsymbol\rho}d^{2}\mathbf{\boldsymbol\rho'}
\exp\Big(-\frac{\sqrt{\mathbf{\boldsymbol\rho}^{2}+r_{0}^{2}}}{b}\Big) \exp\Big(-\frac{\sqrt{\mathbf{\boldsymbol\rho'}^{2}+r_{0}^{2}}}{b}\Big)\exp[-i\beta_{h}\mathbf{q}\boldsymbol\rho]
\exp[-i\beta_{e}\mathbf{q}\boldsymbol\rho'] \int_{0}^{+\infty}d\chi\frac{\chi}{\sqrt{\chi^2+L^2}}\int_{0}^{2\pi}d\phi e^{iq\chi\cos\phi}= \\ &=C\frac{2\pi e^{-qL}}{q}\int d^{2}\mathbf{\boldsymbol\rho}d^{2}\mathbf{\boldsymbol\rho'}
\exp\Big(-\frac{\sqrt{\mathbf{\boldsymbol\rho}^{2}+r_{0}^{2}}}{b}\Big) \exp\Big(-\frac{\sqrt{\mathbf{\boldsymbol\rho'}^{2}+r_{0}^{2}}}{b}\Big)\exp[-i\beta_{h}\mathbf{q}\boldsymbol\rho]
\exp[-i\beta_{e}\mathbf{q}\boldsymbol\rho'].
\label{Ie1h2_2}
\end{align}
\end{widetext}
The same factorization can be done as in the integral (\ref{Ie1e2}) and expression (\ref{Ie1h2_2}) yields
\begin{equation}
\mathcal{I}_{eh'}(\mathbf{q}\rightarrow 0)=\frac{e^2}{2\pi\epsilon\epsilon_{0}A}\cdot\frac{e^{-qL}}{q}
\label{Ie1h2_3}
\end{equation}
One can check that the same expression is valid for $\mathcal{I}_{he'}(\mathbf{q}\rightarrow 0)$. Finally, the sum of four integrals yields
\begin{eqnarray}
V^{q\rightarrow 0}_{dir}=\lim_{q\rightarrow 0}\frac{e^2}{\epsilon\epsilon_{0}A}\frac{(1-e^{-qL})}{q}=\frac{e^2}{\epsilon\epsilon_{0}A}L.
\label{Vdir_fin}
\end{eqnarray}
The similar result was obtained using another approach by authors in Ref. [\onlinecite{de-Leon}].
\\

The exciton exchange interaction $V^{X}_{exch}$ can be written in the same way as direct interaction from the general form (\ref{Vx})
\begin{widetext}
\begin{align*}
&V^{X}_{exch}(\mathbf{Q},\mathbf{Q'},\mathbf{q})=\frac{e^{2}}{4\pi\epsilon\epsilon_{0}A^{2}}\frac{\exp(2r_{0}/b)}{(2\pi b(b+r_{0}))^{2}}\int d^{2}\mathbf{r}_{e}d^{2}\mathbf{r}_{h}d^{2}\mathbf{r}_{e'}d^{2}\mathbf{r}_{h'} \exp\Big(-\frac{\sqrt{(\mathbf{r}_{e}-\mathbf{r}_{h})^2+r_{0}^{2}}}{b}\Big) \exp\Big(-\frac{\sqrt{(\mathbf{r}_{e'}-\mathbf{r}_{h'})^2+r_{0}^{2}}}{b}\Big) \\ &\exp[-i\mathbf{Q}(\beta_{e}\mathbf{r}_{e}+\beta_{h}\mathbf{r}_{h})]\exp[-i\mathbf{Q'}(\beta_{e}\mathbf{r}_{e'}+\beta_{h}\mathbf{r}_{h'})] \Big[\frac{1}{|\mathbf{r}_{e}-\mathbf{r}_{e'}|}+\frac{1}{|\mathbf{r}_{h}-\mathbf{r}_{h'}|}-\frac{1}{\sqrt{(\mathbf{r_{e}}-\mathbf{r_{h'}})^{2}+L^{2}}}-\frac{1}{\sqrt{(\mathbf{r_{h}}-\mathbf{r_{e'}})^{2}+L^{2}}}\Big]\\ &\exp[i(\mathbf{Q}+\mathbf{q})(\beta_{e}\mathbf{r}_{e'}+\beta_{h}\mathbf{r}_{h'})]\exp[i(\mathbf{Q'}-\mathbf{q})(\beta_{e}\mathbf{r}_{e}+\beta_{h}\mathbf{r}_{h})]=\frac{e^{2}}{4\pi\epsilon\epsilon_{0}A^{2}}\frac{\exp(2r_{0}/b)}{(2\pi b(b+r_{0}))^{2}}\int d^{2}\mathbf{\boldsymbol\rho}d^{2}\mathbf{\boldsymbol\rho'}d^{2}\mathbf{R}d^{2}\mathbf{R'} \exp\Big(-\frac{\sqrt{\mathbf{\boldsymbol\rho}^{2}+r_{0}^{2}}}{b}\Big) \\ &\exp\Big(-\frac{\sqrt{\mathbf{\boldsymbol\rho'}^{2}+r_{0}^{2}}}{b}\Big)\exp[-i\Delta\mathbf{Q}(\mathbf{R}-\mathbf{R'})]\exp[-i\mathbf{q}(\mathbf{R}-\mathbf{R'})]\Big[\frac{1}{|\beta_{h}(\mathbf{\boldsymbol\rho}-\mathbf{\boldsymbol\rho'})+\mathbf{R}-\mathbf{R'}|}+\frac{1}{|-\beta_{e}(\mathbf{\boldsymbol\rho}-\mathbf{\boldsymbol\rho'})+\mathbf{R}-\mathbf{R'}|}- \\ &-\frac{1}{\sqrt{(\beta_{h}\mathbf{\boldsymbol\rho}+\beta_{e}\mathbf{\boldsymbol\rho'}+\mathbf{R}-\mathbf{R'})^{2}+L^{2}}}-\frac{1}{\sqrt{(-\beta_{e}\mathbf{\boldsymbol\rho}-\beta_{h}\mathbf{\boldsymbol\rho'}+\mathbf{R}-\mathbf{R'})^{2}+L^{2}}}\Big],
\end{align*}
\end{widetext}
where we defined exchanged momentum between electrons as $\Delta\mathbf{Q}=\mathbf{Q}-\mathbf{Q'}$. One can see that for small exchanged momentum between exciton which is true for weakly interacting exciton gas the expression for $V^{X}_{exch}$ coincides with $V_{dir}$ for $q\rightarrow 0$.\\

\emph{Electron and hole exchange interaction of indirect excitons.}
The general form of electron exchange interaction for indirect excitons is given by Eq. (\ref{Vexch_e}).
For the calculations it is more convenient to choose the indirect exciton wave function in the form (\ref{phi1}). Thus the exchange interaction matrix element yields
\begin{widetext}
\begin{align*}
&V^{e}_{exch}(\mathbf{Q},\mathbf{Q'},\mathbf{q})=-\frac{e^2}{4\pi\epsilon\epsilon_{0}A^2}\frac{4}{\pi^{2}a_{B}^{4}}\int d^{2}\mathbf{r_{e}}d^{2}\mathbf{r_{h}}d^{2}\mathbf{r_{e'}}d^{2}\mathbf{r_{h'}}
\exp\Big(-\frac{\sqrt{(\mathbf{r}_{e}-\mathbf{r}_{h})^{2}+L^{2}}}{a_{B}}\Big)\exp\Big(-\frac{\sqrt{(\mathbf{r}_{e'}-\mathbf{r}_{h'})^{2}+L^{2}}}{a_{B}}\Big)\times \\  &\times \exp\Big(-\frac{\sqrt{(\mathbf{r}_{e'}-\mathbf{r}_{h})^{2}+L^{2}}}{a_{B}}\Big)\exp\Big(-\frac{\sqrt{(\mathbf{r}_{e}-\mathbf{r}_{h'})^{2}+L^{2}}}{a_{B}}\Big)\exp[-i\mathbf{Q}(\beta_{e}\mathbf{r}_{e}+\beta_{h}\mathbf{r}_{h})]\exp[-i\mathbf{Q'}(\beta_{e}\mathbf{r}_{e'}+\beta_{h}\mathbf{r}_{h'})]\times\\ &\exp[i(\mathbf{Q}+\mathbf{q})(\beta_{e}\mathbf{r}_{e}+\beta_{h}\mathbf{r}_{h})]\exp[i(\mathbf{Q'}-\mathbf{q})(\beta_{e}\mathbf{r}_{e'}+\beta_{h}\mathbf{r}_{h'})]\Big[\frac{1}{|\mathbf{r}_{e}-\mathbf{r}_{e'}|}+\frac{1}{|\mathbf{r}_{h}-\mathbf{r}_{h'}|}-\frac{1}{\sqrt{(\mathbf{r}_{e}-\mathbf{r}_{h'})^{2}+L^{2}}}-\frac{1}{\sqrt{(\mathbf{r}_{h}-\mathbf{r}_{e'})^{2}+L^{2}}}\Big].
\end{align*}
\end{widetext}
The exact calculation of exchange integral is straightforward but tedious. Using the same steps as for direct interaction calculation and performing in the end substitutions $\mathbf{y}_{1}=(\boldsymbol\xi-\beta_{e}\boldsymbol\rho-\beta_{h}\boldsymbol\rho')/a_{B}$, $\mathbf{y}_{2}=(\boldsymbol\xi+\beta_{h}\boldsymbol\rho+\beta_{e}\boldsymbol\rho')/a_{B}$, $\mathbf{x}=\boldsymbol\rho/a_{B}$ and $\widetilde{L}=L/a_{B}$ one gets the final expression of electron exchange interaction
\begin{equation}
V^{e}_{exch}=-\frac{e^{2}}{4\pi\epsilon\epsilon_{0}A}\left(\frac{2}{\pi}\right)^{2}a_{B}\cdot\mathcal{I}^{e}_{exch}(\Delta Q,q,\Theta,\beta_{e}),
\label{Vexch_e_fin}
\end{equation}
where the exchange integral is given by
\begin{widetext}
\begin{align}\notag
&\mathcal{I}^{e}_{exch}(\Delta Q,q,\Theta,\beta_{e})=\int_{0}^{\infty}dx\int_{0}^{2\pi}d\Theta_{x}\int_{0}^{\infty}
dy_{1}\int_{0}^{2\pi}d\Theta_{1}\int_{0}^{\infty}dy_{2}\int_{0}^{2\pi}d\Theta_{2}\cdot xy_{1}y_{2}\cos\Big[\Delta Qa_{B}[\beta_{e}x\cos(\Theta-\Theta_{x})+ \\ \notag &+\beta_{e}y_{1}\cos(\Theta-\Theta_{1})]+qa_{B}[-x\cos\Theta_{x}-\beta_{e}y_{1}\cos\Theta_{1}+(1-\beta_{e}y_{2}\cos\Theta_{2})]\Big]\exp(-\sqrt{x^{2}+\widetilde{L}^{2}})
\exp(-\sqrt{y_{1}^{2}+\widetilde{L}^{2}})\times \\ \notag &\times \exp(-\sqrt{y_{2}^{2}+\widetilde{L}^{2}})\exp(-\sqrt{(y_{2}\cos\Theta_{2}-y_{1}\cos\Theta_{1}-x\cos\Theta_{x})^{2}+(y_{2}\sin\Theta_{2}-y_{1}\sin\Theta_{1}-x\sin\Theta_{x})^{2}+\widetilde{L}^{2}})\times\\ &\times\Big[\frac{1}{\sqrt{y_{1}^{2}+x^{2}+2y_{1}x\cos(\Theta_{1}-\Theta_{x})}}
\frac{1}{\sqrt{y_{2}^{2}+x^{2}+2y_{2}x\cos(\Theta_{2}-\Theta_{x})}}-\frac{1}{\sqrt{y_{1}^{2}+\widetilde{L}^{2}}}-\frac{1}{\sqrt{y_{2}^{2}+\widetilde{L}^{2}}}\Big],
\label{Iexch}
\end{align}
\end{widetext}
with $\Theta$ being the angle between $\Delta\mathbf{Q}$ and $\mathbf{q}$. We are interested in the case when $\Delta Q=0$ and $q=0$. The calculation of exchange integral thus requires numerical integration with multidimensional Monte Carlo algorithm. Moreover, it is obvious that similarly to the case of direct excitons electron and holes exchange interactions have the same value for $q\rightarrow 0$.

\end{document}